\documentclass[11pt]{article}
\usepackage{fleqn,cospar,epsfig}

\usepackage{url}

\usepackage{graphicx}
\def\eps@scaling{.95}%
\newcommand\epsscale[1]{\gdef\eps@scaling{#1}}%
\newcommand\plotone[1]{%
 \centering
 \leavevmode
 \includegraphics[natwidth={\eps@scaling\columnwidth}]{#1}%
}%
\newcommand\plottwo[2]{%
 \centering
 \leavevmode
 \columnwidth=.45\columnwidth
 \includegraphics[natwidth={\eps@scaling\columnwidth}]{#1}%
 \hfil
 \includegraphics[natwidth={\eps@scaling\columnwidth}]{#2}%
}%

\usepackage[figuresright]{rotating}
\newcommand\tablerefs[1]{\appgdef\tblnote@list{\@tableref{#1}}}%
\def\@tableref#1{%
 \par
 \vspace*{3ex}%
 {\parbox{\pt@width}{\hskip1em\rmfamily References. --- #1}\par}%
}%
\newcommand\tablecomments[1]{\appgdef\tblnote@list{\@tablecom{#1}}}%
\def\@tablecom#1{%
 \par
 \vspace*{3ex}%
 {\parbox{\pt@width}{\hskip1em\rmfamily Note. --- #1}\par}%
}%
\def\@tableref@pptt#1{%
 \par
 \vspace*{3ex}{%
 \parbox{\pt@width}{\hskip1em{\sc References.---}#1}\par}%
}%
\def\@tablecom@pptt#1{%
 \vspace*{3ex}{%
 \parbox{\pt@width}{\hskip1em{\sc Note.---}#1}\par}%
}%

\title{Oscillations During Thermonuclear X-ray Bursts}

\author{Tod E. Strohmayer\address{NASA's Goddard Space Flight Center, 
    Mail Code 662, Greenbelt, MD 20771, USA}}

\begin{document}

\maketitle

\begin{abstract}

High amplitude, nearly coherent X-ray brightness oscillations during 
thermonuclear X-ray bursts were discovered with the {\it Rossi} X-ray Timing 
Explorer (RXTE) in early 1996. Spectral and timing evidence strongly 
supports the conclusion that these oscillations are caused by rotational 
modulation of the burst emission and that they reveal the spin frequency of
neutron stars in low mass X-ray binaries, a long sought goal of X-ray
astronomy. Studies carried out over the past year have led to the 
discovery of burst oscillations in four new sources, bringing to ten the 
number with confirmed burst oscillations. I review the status of our knowledge
of these oscillations and indicate how they can be used to probe the physics 
of neutron stars. For a few burst oscillation sources it has been proposed 
that the strongest and most ubiquitous frequency is actually the first overtone
of the spin frequency and hence that two nearly antipodal hot spots are present
on the neutron star. This inference has important implications for both the 
physics of thermonuclear burning as well as the mass - radius relation for 
neutron stars, so its confirmation is crucial. I discuss recent attempts to 
confirm this hypothesis for 4U 1636-53, the source for which a signal at the 
putative fundamental (290 Hz) has been claimed.

\end{abstract}

\section*{INTRODUCTION}

Since its launch in December, 1995 NASA's {\it Rossi} X-ray Timing Explorer has
provided astronomers with a fundamental new view of neutron stars, and in 
particular those which are accreting in binary systems. Since the discovery of
rapidly rotating neutron stars as millisecond radio pulsars it has been 
suspected that neutron stars in low mass X-ray binaries (LMXB) are spun up to 
millisecond periods by the capture of angular momentum via mass transfer
from an accretion disk. Efforts to confirm this hypothesis by detecting
millisecond X-ray pulsars in LMXBs went unrewarded for many years. This 
situation changed dramatically with the advent of RXTE. Within a few months of 
its launch RXTE observations had provided strong evidence suggesting that 
neutron stars in LMXB are spinning with frequencies $\ge 300$ Hz. These first 
indications came with the discovery of high frequency (millisecond) X-ray 
brightness oscillations, ``burst oscillations,'' during thermonuclear (Type I) 
X-ray bursts from several neutron star LMXB systems (see Strohmayer et al. 
1996; Smith, Morgan \& Bradt 1997; Zhang et al. 1996). 

As of this writing burst oscillation detections have been claimed for a total 
of ten different LMXB systems, with four of these only appearing in the past
few months (see Strohmayer et al. 1996; Zhang et al. 1996; Smith, Morgan \& 
Bradt 1997; Strohmayer et al. 1997; Markwardt, Strohmayer, \& Swank 1998;
Zhang et al. 1998; Wijnands, Strohmayer \& Franco 2000; Galloway et al. 2000;
Chakrabarty 2000; Heise 2000). The (1) observed amplitudes, (2) high coherence,
and (3) long-term stability of these oscillations provide overwhelming 
evidence that they are produced by spin modulation of the X-ray burst flux 
coming from the neutron star surface. Moreover, there properties are 
dramatically in contradiction to what would be expected from, e.g., models 
where the modulations reside in the accretion disk surrounding the neutron 
star. 

The observed frequencies are in 
the range from $\approx 270 - 620$ Hz, bearing a strong similarity to the 
observed frequency distribution of binary millisecond radio pulsars 
(Taylor, Manchester \& Lyne 1993; Bildsten 2000) and consistent with some 
theoretical determinations of spin periods which can be reached via 
accretion-induced spin-up (Webbink, Rappaport \& Savonije 1983). Very recently
Heise et al. (2000) reported evidence of a burst oscillation from the 
only known accreting millisecond pulsar SAX J1808-369 at a frequency 
consistent with its precisely known spin frequency of 401 Hz (Wijnands \&
van der Klis 1998; Chakrabarty \& Morgan 1998). This provides 
additional compelling evidence that the burst oscillations are produced by
spin modulation of the X-ray burst flux.
 
In this contribution I will briefly review our observational 
understanding of these oscillations, with emphasis on how they can be 
understood in the context of spin modulation of the X-ray burst flux. As this
field is evolving rapidly what I present will necessarily be a snapshot of our 
current understanding. I will discuss how detailed modelling of the 
oscillations can be used to place interesting constraints on the masses and 
radii of neutron stars and therefore the equation of state (EOS) of 
supranuclear density matter. Inferences which can be drawn regarding the 
physics of thermonuclear burning will also be discussed. I will show that the 
physical inferences which can be drawn in these areas depend crucially on 
whether one or two (anitpodal) hot spots produce the observed modulations. I 
will present an analysis to try and confirm the two spot scenario for 
4U 1636-53, the source for which the most compelling case for two spots has 
been made (Miller 1999). I conclude with some outstanding theoretical 
questions and uncertainties and where future observations and theoretical work 
may lead.

\section*{THEORETICAL EXPECTATIONS}

The notion that X-ray bursts are casued by thermonuclear instabilities in the
accreted layer on the surface of a neutron star is now universally 
accepted. There is no doubt that interesting puzzles remain and our detailed 
understanding is incomplete, but the basic model is firmly established.
The thermonuclear instability which triggers an X-ray burst burns in a few 
seconds the fuel which has been accumulated on the surface 
over several hours. This makes it very unlikely that the conditions 
required to trigger the instability will be achieved simultaneously over the 
entire stellar surface. This notion, first emphasized by Joss (1978), 
led to the study of lateral propagation of the burning front over the neutron 
star surface (see Fryxell \& Woosley 1982, Nozakura, Ikeuchi \& Fujimoto 1984, 
and Bildsten 1995). The short risetimes of thermonuclear X-ray bursts 
suggest that convection plays an important role in the physics of the burning 
front propagation, especially in the low accretion rate regime which leads to 
large ignition columns (see Bildsten 1998 for a recent review of
thermonuclear burning on neutron stars). These 
studies emphasized that the physics of thermonuclear burning is necessarily a 
multi-dimensional problem and that {\it localized} burning is to be expected, 
especially at the onset of bursts.  As I will describe below, the properties of
oscillations near burst onset described above fit nicely with this picture of 
thermonuclear burning on neutron stars.

\section*{PROPERTIES OF BURST OSCILLATIONS}

Burst oscillations with a frequency of 363 Hz were first discovered from the 
LMXB 4U 1728-34 by Strohmayer et al. (1996). Since then oscillations in 
an additional nine sources have been reported. The sources and their 
observed frequencies are given in Table 1. In the remainder of this section I 
will briefly review the important observational properties of these 
oscillations and lay out the evidence supporting spin modulation as the 
mechanism.

\begin{table}[tbp]
%
\begin{center}
\centerline{{\bf Table 1.} Burst Oscillations Sources and Properties}
\begin{tabular}{cccc}
\hline
\\
Sources & Frequency (Hz) & $\Delta\nu$ (kHz QPO, in Hz) & References$^1$ \cr
\hline
4U 1728-34 & 363 & 363 - 280 & 1, 2, 3, 4, 5, 13, 14 \cr
4U 1636-53 & 290, 580 & 251 & 6, 7 \cr
4U 1702-429 & 330 & 315 - 344 & 4, 9 \cr
KS 1731-260 & 524 & 260 & 10, 11, 12 \cr
Galactic Center & 589 & Unknown & 15 \cr
Aql X-1 & 549 & Unknown & 16, 17 \cr
X1658-298 & 567 & Unknown & 18 \cr
4U 1916-053 & 270 & 290 - 348 & 19, 20 \cr
4U 1608-52 & 619 & 225 - 325 & 8, 21 \cr
SAX J1808-369 & 401 & Unknown & 22,23 \cr
\hline
\end{tabular}
\end{center}
\hfil\hspace{\fill}

$^1$References: (1) Strohmayer et al. (1996); (2) Strohmayer,
Zhang, \& Swank (1997); (3) Mendez \& van der Klis (1999); (4) Strohmayer 
\& Markwardt (1999); (5) Strohmayer et al. (1998b); (6) Strohmayer et al. 
(1998a); (7) Miller (1999); (8) Mendez et al. (1998); (9) Markwardt, 
Strohmayer
\& Swank (1999) (10) Smith, Morgan, \& Bradt (1997); (11) Wijnands \& van der
Klis (1997); (12) Muno et al. (2000); (13) van Straaten et al. (2000); (14)
Franco (2000); (15) Strohmayer et al (1997); (16) Zhang et al. (1998); (17)
Ford (1999); (18) Wijnands, Strohmayer \& Franco (2000); (19) Boirin et al. 
(2000); (20) Galloway et al. (2000); (21) Chakrabarty (2000); (22) Heise 
(2000); (23) Ford (2000)

\end{table} 

\subsection*{Oscillations at Burst Onset}

Some bursts show very strong oscillations during the $\approx 1 - 2$ s 
rises typical of thermonuclear bursts. Strohmayer, Zhang \& Swank (1997)
showed that some bursts from 4U 1728-34 have oscillation amplitudes as large as
43\% within 0.1 s of the observed onset of the burst. They also showed that 
the oscillation amplitude decreased monotonically as the burst flux increased 
during the rising portion of the burst lightcurve. Strohmayer et al (1998a)
reported on strong pulsations in 4U 1636-53 at 580 Hz with an amplitude of 
$\approx 75$\% only $\sim 0.1$ s after detection of burst onset. In both of 
these studies the quoted amplitude is defined as half the peak to peak 
intensity modulation divided by the mean intensity. The mean level is 
determined by first subtracting the pre-burst intensity.  
The presence of modulations of the thermal burst flux 
approaching 100\% right at burst onset fits nicely with the idea that early 
in the burst there exists a localized hot spot which is then modulated by the 
spin of the neutron star. In this scenario the largest modulation amplitudes 
are produced when the spot is smallest, as the spot grows to encompass more of 
the neutron star surface, the amplitude drops, consistent with the 
observations. X-ray spectroscopy during burst rise also supports the inference 
that X-ray emission is localized on the neutron star near the onset of bursts.
Strohmayer, Zhang, \& Swank (1997) found that during burst rise the flux is
underluminous compared with intervals later in the burst which have the same
observed black body temperature, suggesting that during the rise only a 
portion of the surface of the neutron star is producing the X-ray emission. 
As the burst progresses the burning area increases in size until the entire 
surface is involved.

\subsection*{Coherence of Burst Oscillations}

The observed oscillation frequency during a burst is usually not constant.
Often the frequency is observed to increase by $\approx 1 - 3$ Hz in the 
cooling tail, reaching a plateau or asymptotic limit (see Strohmayer et al. 
1998a). This behavior is common to all the burst oscillation sources, and it
would appear that the same physical mechanism is involved, however, there
have been reports of decreases in the oscillation frequency with time. For
example, Strohmayer (1999) and Miller (1999) identified a burst from 4U 1636-53
with a spin down of the oscillations in the decaying tail. This burst also had
an unusually long decaying tail which may have been related to a ``reheating''
episode and could also acount for the spin down.
Muno et al. (2000) reported on a burst from KS 1731-260 which also
showed an episode where the frequency dropped, however, in this case they found
no evidence for unusual flux enhancements or spectral varations during the 
episode. Strohmayer (1997) suggested that relative motions of the hot spot 
and/or burning front on the neutron star surface might also introduce both 
spin up and spin down episodes. 
 
Strohmayer et. al (1997) have argued that the time evolution of the frequency 
results from angular momentum conservation of the thermonuclear shell. 
The burst expands the shell, increasing its rotational moment of inertia and 
slowing its spin rate. Near burst onset the shell is thickest and thus the 
observed frequency lowest. The shell spins back up as it cools and 
recouples to the underlying neutron star. Calculations indicate that the 
$\sim 10$ m thick pre-burst shell expands to $\sim 30$ m during the flash 
(see Joss 1978; Bildsten 1995; Cumming \& Bildsten 2000), which gives a 
frequency shift of $\approx 2 \ \nu_{spin} (20 \ {\rm m}/ R)$,
where $\nu_{spin}$ and $R$ are the stellar spin frequency and radius, 
respectively. For the several hundred Hz spin frequencies inferred from 
burst oscillations this gives a shift of $\sim 2$ Hz, similar to that observed.
Strohmayer \& Markwardt (1999) showed that the frequency evolution in 
4U 1728-34 and 4U 1702-429 is highly phase coherent. They modelled the 
frequency drift and showed that a simple exponential ``chirp" model of the 
form $\nu (t) = \nu_0 (1 - \delta_{\nu} \exp(-t/\tau) )$, works remarkably 
well, producing quality factors $Q \equiv \nu_0 / \Delta\nu_{FWHM} \sim 4,000$.
Muno et al. (2000) performed a similar analysis on bursts from KS 1731-26 and
concluded that the burst oscillations from this source were also phase 
coherent. These results argue strongly that the mechanism which produces the 
modulations is intrinsically a highly coherent one. 

Recently, Galloway et al. (2000) reported a 3.5 Hz frequency shift in a burst 
from 4U 1916-053 with 272 Hz oscillations. They suggested that such a large
change, $\sim 1.3 \%$ might be inconsistent with expansion of the 
thermonuclear burning layer because of the magnitude of the implied height
change of $\sim 80$ m. Wijnands, Strohmayer \& Franco (2000) found a $\sim 5$ 
Hz frequency shift in a burst from 4U 1658-298 with a 567 Hz oscillation, which
may also be uncomfortably large given current estimate of the expansion of
burning layers (Cumming \& Bildsten 2000). Note, however, that the current 
theoretical estimates do not include the rotational lowering of the effective 
surface gravity and are also hydrostatic calculations. Dynamic motions of the 
layer may also contribute to changes in the height of the burning layer. These 
effects could increase the height of the burning layer and allow for greater 
frequency drifts than current calculations suggest. Also, a combination of 
height changes as well as lateral motions with respect to the neutron star 
surface may be at work simultaneously. Clearly more theoretical work is
required to determine if expansion of the burning layer is indeed the 
primary mechanism responsible for the frequency drifts and also to understand 
the nature of the less commonly observed spin down episodes.  

\subsection*{Long Term Stability}

The accretion-induced rate of change of the neutron star spin frequency in a 
LMXB is approximately $1.8 \times 10^{-6}$ Hz yr$^{-1}$ for typical neutron
star and LMXB parameters.  The Doppler shift due to orbital motion of the
binary can produce a frequency shift of magnitude
$ \Delta\nu / \nu = v \sin i /c \approx  2.05 \times 10^{-3}$, again for
representative LMXB system parameters. This doppler shift easily dominates 
over any possible accretion-induced spin change on orbital to several year 
timescales. Therefore the extent to which the observed burst oscillation 
frequencies are consistent with possible orbital Doppler shifts, but 
otherwise stable over $\approx$ year timescales, provides strong support 
for a highly coherent mechanism which sets the observed frequency. 

At present, the best source available to study the long term stability of burst
oscillations is 4U 1728-34. Strohmayer et al. (1998b) compared the observed
asymptotic frequencies in the decaying tails of bursts separated in time
by $\approx 1.6$ years. They found the burst frequency to be highly stable,
with an estimated time scale to change the oscillation period of 
about 23,000 year. It was also suggested that the stability of the 
asymptotic periods might be used to infer the X-ray mass function of
LMXB by comparing the observed asymptotic period distribution of many
bursts and searching for an orbital Doppler shift. The source 4U 1636-53 is a 
good candidate for such an effort because its orbital period is known (3.8
hrs). Strohmayer et al. (1998b) compared the highest observed frequencies in 
three different bursts from 4U 1636-53. The frequencies in these bursts alone 
were consistent with a typical orbital velocity for the neutron star. However, 
study of additional bursts reveals a greater range of highest 
frequencies than can likely be accounted for by orbital motion alone (Giles \&
Strohmayer 2001). A possible explanation of this within the context of
the spin modulation scenario is that not every burst has relaxed to the 
asymptotic value before the oscillations fade below the detection level. 
Nevertheless, the observed distribution of frequencies in 4U 1636-53 does 
suggest the existence of an upper limit, which can naturally be associated 
with the spin frequency (Giles \& Strohmayer 2001).

\subsection*{Burst Oscillations and Source State}

Several recent studies have focused on how the presence and properties of
burst oscillations may or may not correlate with other properties of these 
sources, for example, their spectral state and mass accretion rate. 
Muno et al. (2000) found that
bursts from KS 1731-26 with oscillations appear to only occur when the source
is on the banana branch in the X-ray color-color diagram. They also found that
these bursts were all radius expansion bursts. Cumming \& Bildsten (2000) 
suggested that such bursts were likely pure Helium flashes and that it would be
more likely for these to show oscillations because the radiative diffusion time
is short compared to the inferred shearing time of the burning layer, 
making it more likely that a modulation would survive. Franco (2000) and 
van Straaten et al. (2000) showed that bursts from 4U 1728-34 with oscillations
also occur preferentially on the banana branch, but they did not find a 
similar relationship with radius expansion as for KS 1731-26. Franco (2000) 
also found that the strength of oscillations was correlated with position in 
the color-color diagram. These results are begining to provide new insights 
into how mass accretion rate effects thermonuclear burning on neutron stars.

\section*{BURST OSCILLATIONS AS PROBES OF NEUTRON STARS}

Precise modelling of burst oscillations holds great promise for 
providing new insights into a variety of physics questions concerning the 
structure and evolution of neutron stars. For example, pulse amplitudes and
shapes produced by rotational modulation of a hot spot contain information on
both the compactness, $M/R$, of the neutron star as well as the rotational 
velocity of the hot spot. Since the spin frequency is known the velocity is 
directly proportional to the stellar radius $R$. In addition, the rotational
motion provides us with a snapshot view of the propagation of the thermonuclear
instability.

\subsection*{M - R Constraints}

One of the crucial bits of physics that makes compactness constraints possible 
is the bending of photon trajectories in a strong gravitational field. The 
strength of such deflection is a function of the stellar compactness, 
$GM/c^2 R$, with more compact stars producing greater deflections and therefore
weaker spin modulations. An upper limit on the compactness can be set since a 
star more compact than this limit would not be able to produce a modulation as 
large as observed. Complementary information comes from the pulse shape, which 
can be inferred from the strength of harmonics. Information on both the 
amplitude and harmonic content can thus be used to bound the compactness. 

Stellar rotation also plays a role in the observed properties
of spin modulation pulsations. For example, a 10 km radius neutron star 
spinning at 400 Hz has a surface velocity of $v_{spin}/c \le 2\pi
\nu_{spin} R/c \approx 0.084$ at the rotational equator. This motion produces 
a relativistic abberation as well as a Doppler shift of magnitude $\Delta E / E
\approx v_{spin}/c$ (see Chen \& Shaham 1989). Measurement of the pulse phase 
dependent Doppler shift in the X-ray spectrum would provide additional evidence
supporting the spin modulation model and also provides one of the few direct
methods to infer the radius of a neutron star. Ford (1999) has analysed data 
during a burst from Aql X-1 and finds that the softer photons lag higher 
energy photons in a manner which is qualitatively similar to that expected 
from a rotating hot spot.

Detailed modelling of pulse shapes and pulse phase spectra can then
be used to determine a confidence region in the mass - radius plane for neutron
stars. Miller \& Lamb (1998) have investigated the amplitude of rotational
modulation pulsations as well as harmonic content assuming emission from a 
point-like hot spot. They also showed that knowledge of the angular and 
spectral dependence of the emissivity from the neutron star surface can have 
important consequences for the derived constraints. Recently, Nath, 
Strohmayer \& Swank (2000) have used a rotating hot spot model to fit 
bolometric 
pulse profiles observed on the rising edge of bursts from 4U 1636-53. They 
find that the inferred constraints depend very sensitively on whether or not 
two spots are present. Much more restrictive compactness constraints can be 
achieved if the two spot hypothesis is correct. The main reason for this being
that large amplitudes are much more difficult to achieve with two spots 
than one. The large observed amplitudes from 4U 1636-53 require a large neutron
star which in turn requires a very stiff EOS. The inferred radius is also 
sensitive to this two spot assumption, since uncertainty in the number of spots
gives a factor of two ambiguity in the rotational velocity. 

\subsection*{Physics of Nuclear Burning and Propagation}

The properties of burst oscillations can tell us a great deal about the 
processes of nuclear burning on neutron stars. The amplitude evolution
during the rising phase of bursts contains information on how rapidly
the flame front is propagating. If the anitpodal spot hypothesis to explain
the presence of a subharmonic in 4U 1636-53 is correct, then it has important
implications for the propagation of the instability from one pole to
another in $\approx$ 0.2 s (see Miller 1999). In addition, a two pole flux 
anistropy suggests that the nuclear fuel is likely pooled by some mechanism, 
perhaps associated with the magnetic field of the star. In order to have a 
robust understanding of the implications of burst oscillations for neutron star
structure and thermonuclear propagation it is essential to remove any 
ambiguity concerning the number of hot spots producing the observed 
modulations.

\section*{4U 1636-53: WHAT IS THE SPIN FREQUENCY?}

Zhang et al. (1996) reported the discovery of 580 Hz oscillations in bursts
from 4U 1636-53. Observations of the accretion driven flux from this source
yielded a pair of kilo-Hertz quasiperiodic oscillations (QPO) with a frequency
separation of about 280 Hz (see Mendez et al. 1998). The beat frequency 
interpretation for the twin kHz peaks suggests that the observed frequency
separation of the kHz QPO peaks should be close to the spin frequency of the 
neutron star (see Strohmayer et al. 1996; Miller, Lamb \& Psaltis 1998). 
The observed $\sim 280$ Hz kHz QPO separation in 4U 1636-53 prompted Miller 
(1999) to search for a $\sim 290$ Hz subharmonic of the 580 Hz signal during 
bursts from this source. Miller (1999) added the signals from 5 different 
bursts and reported a detection of the subharmonic with a chance detection 
probability of $4 \times 10^{-5}$ and suggested that 290 Hz is the neutron star
spin frequency.  

\subsection*{Coherent Addition of Signals from Different Bursts}

Miller (1999) used a ``matched waveform filtering'' or ``template filtering''
technique to search for the subharmonic of the already detected 580 Hz signal
in bursts from 4U 1636-53. He did this by using the observed signal at 580 Hz 
to predict what the 290 Hz signal would be and to then use that as the 
template in his search. Strohmayer \& Markwardt (1999) have investigated the 
phase and frequency evolution of pulsations in bursts using the $Z_n^2$ 
statistic,
\begin{equation}
\hskip 120pt Z_n^2 = 2/N \sum_{k=1}^{n} \left ( \sum_{j=1}^N \cos (k\phi_j)
\right )^2 + \left ( \sum_{j=1}^N \sin (k\phi_j)\right )^2 \; ,
\end{equation}
where $N$ is the total number of photons in the time series, $\phi_j$ are the
phases of each photon derived from a frequency model, $\nu (t)$, {\it vis.}
$\phi_j = 2\pi \int_0^{t_j} \nu (t') dt'$, and $n$ is the total number of
harmonics added together. Here I use this method to search for harmonics and 
the subharmonic of the 580 Hz signal from 4U 1636-53 in a different set of 
bursts than studied by Miller (1999). 

For the burst oscillations, which are highly sinusoidal, it is appropriate to 
set $n=1$. The $Z_1^2$ statistic is particularly well suited to event mode 
data, since no binning is introduced, and it has the same statistical 
properties as the well known Leahy normalized power spectrum, which for a 
Poisson process is distributed as the $\chi^2$ function with 2 degrees of 
freedom. All of the bursts discussed here were observed with the Proportional 
Counter Array (PCA) onboard RXTE and sampled with 125 $\mu$s (1/8192 s) 
resolution. Following Miller (1999) I search for the subharmonic during 
pulsations on the rising edge of bursts. 
Since I am trying to confirm a previous detection I use a different sample 
of bursts than were analysed by Miller (1999). After compiling a catalog of 
all available bursts from 4U 1636-53 I searched for those with significant 
oscillations on the rising edge and found six suitable for the search. 
The bursts I used are summarized in Table 2. 

\begin{table}[tbp]
%
\begin{center}
\centerline{{\bf Table 2.} Bursts from 4U 1636-53 used in the subharmonic 
search}
\begin{tabular}{cccccc}
\hline
\\
Segment & Burst & Obs. ID. & Date & HJD - 2450000$^1$ & $Z_1^2$ \cr
\hline
1 & 1 & 30053-02-02-02 & 8/19/98 & 1044.991053 & 124 \cr
2 & 2 & 30053-02-01-02 & 8/20/98 & 1045.654542 & 54 \cr
3 & 3 & 30053-02-02-00 & 8/20/98 & 1045.719849 & 148 \cr
4 & 4 & 40028-01-02-00 & 2/27/99 & 1236.865609 & 61$^a$ \cr
5 & 4 & 40028-01-02-00 & 2/27/99 & 1236.865609 & 52$^b$ \cr
6 & 5 & 40028-01-08-00 & 6/18/99 & 1348.493173 & 111 \cr
7 & 6 & 40028-01-10-00 & 9/25/99 & 1447.360320 & 69 \cr
\hline
\end{tabular}

\end{center}
\hfil\hspace{\fill}

$^1$Heliocentric Julian Date 

\end{table}

\subsection*{Burst Oscillations in the Complex Plane}

The $\sin\phi$ and $\cos\phi$ terms in the $Z_1^2$ function can be thought of 
as the real and imaginary components of a complex expression, $Z = \sum_j 
( \cos\phi_j + i \sin\phi_j )$. $Z_1^2$ is just the summed squared modulus of 
this expression. Graphically we can represent the evolution of oscillations in 
a given data segment by plotting the positions in the complex plane of the
set of cumulative $Z$s for each segment, with each point being obtained by
adding in the next event in the sum. In such a representation, a coherent 
signal will be evident as a straight line in the complex plane with a 
position angle which represents the phase of the oscillations. The length 
squared of such a line is just the peak $Z_1^2$ for that set of events.

\subsubsection*{Evidence for a Phase Jump}

For each burst in our sample I used frequency models, $\nu (t)$, with both 
a constant and a linearly evolving frequency. I found no significant evidence 
for a changing frequency with one exception. Burst 4 showed an unusual phase 
evolution when the cumulative phases for the best constant frequency model 
were plotted in the complex plane. 
This behavior is illustrated in Figure 1. This burst has a very strong 
oscillation begining $\le 0.05$ s after onset (see Figure 1, left panel). 
The oscillation fades for about 0.1 s, evidenced by the drop in $Z_1^2$, before
strengthening again, however, when the oscillation is seen again there is an 
apparent phase offset compared to what would be expected from the best 
frequency inferred from the first 0.15 s of the burst. The phase offset is 
about 1/4 of a cycle and it was accumulated over $\sim 0.1$ s (see Figure 1, 
right panel), implying a limit on $d P / dt \sim 4 \times
10^{-3}$ s s$^{-1}$ during this time. 
Because of this phase evolution, for the purposes of conducting the 
subharmonic search burst 4 was broken up into two separate segments. Each 
segment was modelled separately using a constant frequency model.

\begin{figure}
\centering
\leavevmode
\includegraphics[width=85mm]{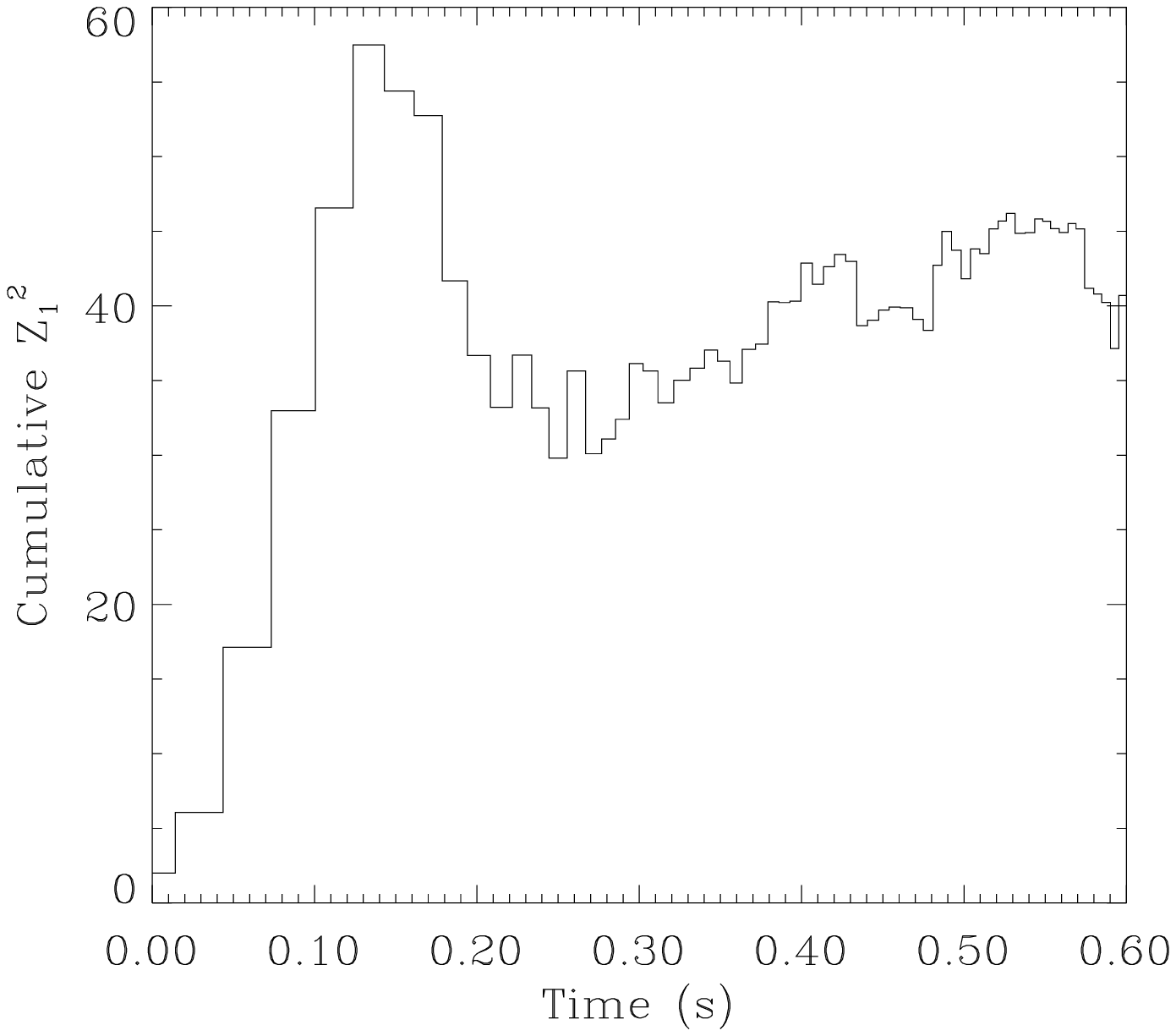}
\includegraphics[width=85mm]{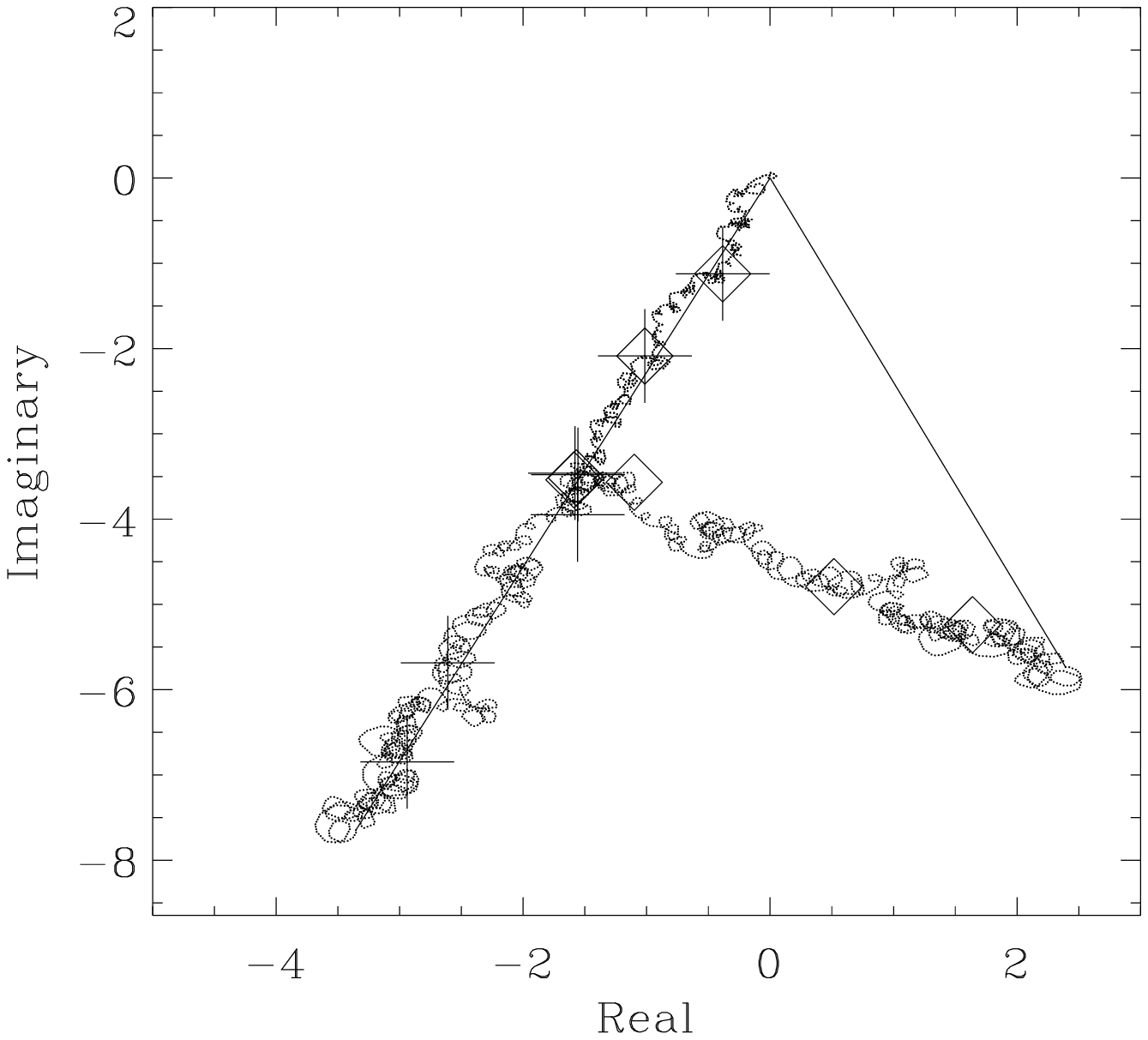}
\caption{Cumulative $Z_1^2$ for burst 4 using a constant frequency of 
580.8 Hz (left panel). There is a very strong oscillation in the first 0.1 s 
of the burst followed by a fading of the oscillation (evidenced by the drop in
$Z_1^2$ between 0.15 and 0.25 s). The signal strengthens again, but there
is a phase offset compared with an extrapolation of the best frequency
inferred from the first 0.15 s of data. The right panel shows the phase 
evolution in the complex plane using the best frequency for the interval from 
0 to 0.15 s. The phase evolution is shown both with (crosses) and without 
(diamonds) a phase shift in the 2nd data segment of $\sim 0.234$ of a cycle . 
Including the phase shift straightens out the bend in the phase evolution. 
The symbols are plotted every 0.05 s.}
\end{figure}

\subsubsection*{Subharmonic Search}

The search procedure had the following steps. 1) I 
first extracted a set of X-ray events from the start of each burst, $t_0$ to 
the peak, $t_{peak}$.
2) I then maximized the $Z_1^2$ signal at 580 Hz separately for each burst. 
I did this by finding the event time $t^i_{max}$ between $t^i_0$ and 
$t^i_{peak}$ which gives the maximum $Z_1^2$. I found, with one exception, 
that for the rising intervals a constant frequency model was adequate. That 
is, I did not find  that a linear time dependence (either increasing or 
decreasing) of the frequency improved $Z_1^2$ significantly, except in the 
one case noted above. Therefore for each data segment I determined the 
constant frequency $\nu ^i$ which gave the maximum $Z_1^2$ for each burst. 
I then used the events and the 
frequency, $\nu ^i$, to compute a set of phases, $\phi^i_j = 2\pi \nu ^i
t_j$, for each burst, where the $i$ subscript identifies the different bursts.
3) I next combined the phases from each burst to determine a maximum $Z_1^2$.
This was done by first determining a phase offset, $\delta\phi_i$, for each 
burst which maximized the $Z_1^2$ signal computed using events from that burst 
{\it and} the first burst, whose phase was used as an initial reference 
without loss of generality. This amounted to computing the following;
\begin{equation}
\hskip 65pt Z_{sum}^2 = {2\over{\sum_{i=1}^{m_{burst}}} N_i} \left [ \left ( 
\sum_{i=1}^{m_{burst}} 
\sum_{j=1}^{N_{i}} \cos (\phi_j^i + \delta\phi^i ) \right )^2 + 
\left ( \sum_{i=1}^{m_{burst}} \sum_{j=1}^{N_{i}} \sin (\phi_j^i + 
\delta\phi^i ) \right )^2 \right ] ,
\end{equation}
where $N_i$ is the number of X-ray events in burst $i$, and $m_{burst}$ is
the number of different bursts added together. The $\delta\phi$ are the 
phase shifts which align each burst with respect to burst 1 (ie. 
$\delta\phi_1 \equiv 0$). 

I carried out this procedure for the 7 data segments in Table 1. The results
are summarized in Figure 2 where I have plotted the cumulative phases for 
each data segment as well as the total coherent sum. The segments are numbered
according to the label in Table 2. The fact that each
line segment is straight to within statistical precision attests to its
phase coherence. The total $Z_{sum}^2$ power at 580 Hz of $\sim 585$ can be 
compared to a power of $\sim 138$ quoted by Miller (1999) for the coherent 
sum of the 5 different bursts he analysed.

\begin{figure}
\centerline{\includegraphics[width=125mm,height=70mm]{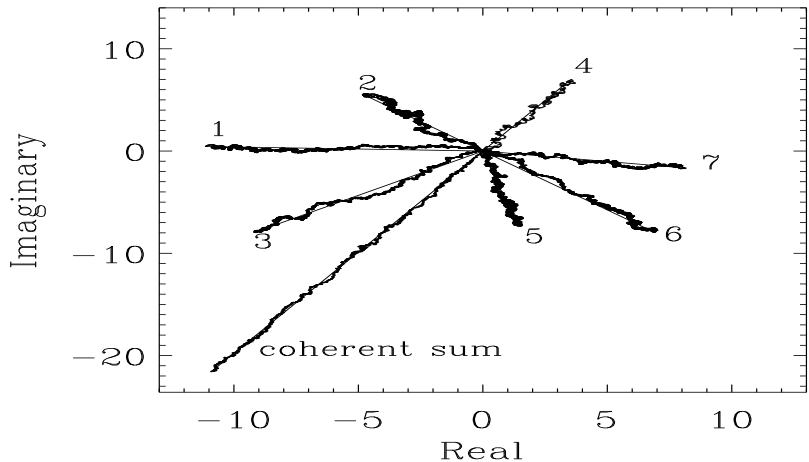}}
\caption{Cumulative phases in the complex plane using the constant frequency 
model for each data segment used in our analysis. The coherent sum of all 7 
intervals is also shown. The fact that each segment can be approximated by a 
straight line, to within statistical precision, is a demonstration that each 
segment is phase coherent. The total $Z_{sum}^2$ at 580 Hz is $\sim 585$. 
This can be compared with a signal of $\sim 138$ obtained by Miller (1999) 
using 5 different bursts. For clarity an arbitrary phase offset has been 
added to some of the segments so they do not overlap.}
\end{figure}

Having coherently added all the data segments I can now conduct harmonic and
subharmonic searches. A harmonic search is straightforward. I simply
evaluated $Z_{sum}^2 (k\phi^i_j)$ with $k=2,3,4 ...$ etc. For any $k$ the 
$Z_{sum}^2$ will be distributed as $\chi^2$ with 2 degrees of freedom, so that
I can determine a probability that a harmonic signal is present by computing
the probability of drawing $Z_{sum}^2 (k\phi^i_j)$ from the $\chi^2$ 
distribution. Evaluating $Z_{sum}^2 (k\phi^i_j)$ for $k=1,2,3,4,5$ did not
give a significant detection of any harmonic. Indeed the pulsations are 
highly sinusoidal, with a maximum power at the 1st harmonic $(k=2)$ of only
2.81. This means that any signal at the first harmonic is at least 15 times
weaker than the 580 Hz signal. 

To search for the subharmonic is a little more difficult because there is a 
$\pi$ phase ambiguity between the crests of any putative 290 Hz oscillation
when combining two different bursts (see Miller 1999). I simply 
multiplied each set of phases by 1/2 to search the first subharmonic, $590/2$
Hz, however, to combine 7 separate segments I must allow for a 
total of $2^6$ = 64 different combinations of the phases. This is equivalent to
a search with 64 independent trials. The results of this search are
shown in Figure 3. I did not detect the subharmonic. I found a peak power at 
the subharmonic of only $\sim 10$, which means that
the signal amplitude at 290 Hz is less than $\sim 13 \%$ of the amplitude at
580 Hz. If the subharmonic signal would have been at the same strength as 
quoted by Miller (1999) I would have found a power at the subharmonic of 
$\sim 110$ which would have been easily detected in these data. Analysis of 
these six bursts using the waveform matching technique of Miller (1999) 
confirms the lack of a significant signal at 290 Hz (C. Miller, personal 
communication).

\begin{figure}
\centering
\leavevmode
\includegraphics[width=85mm]{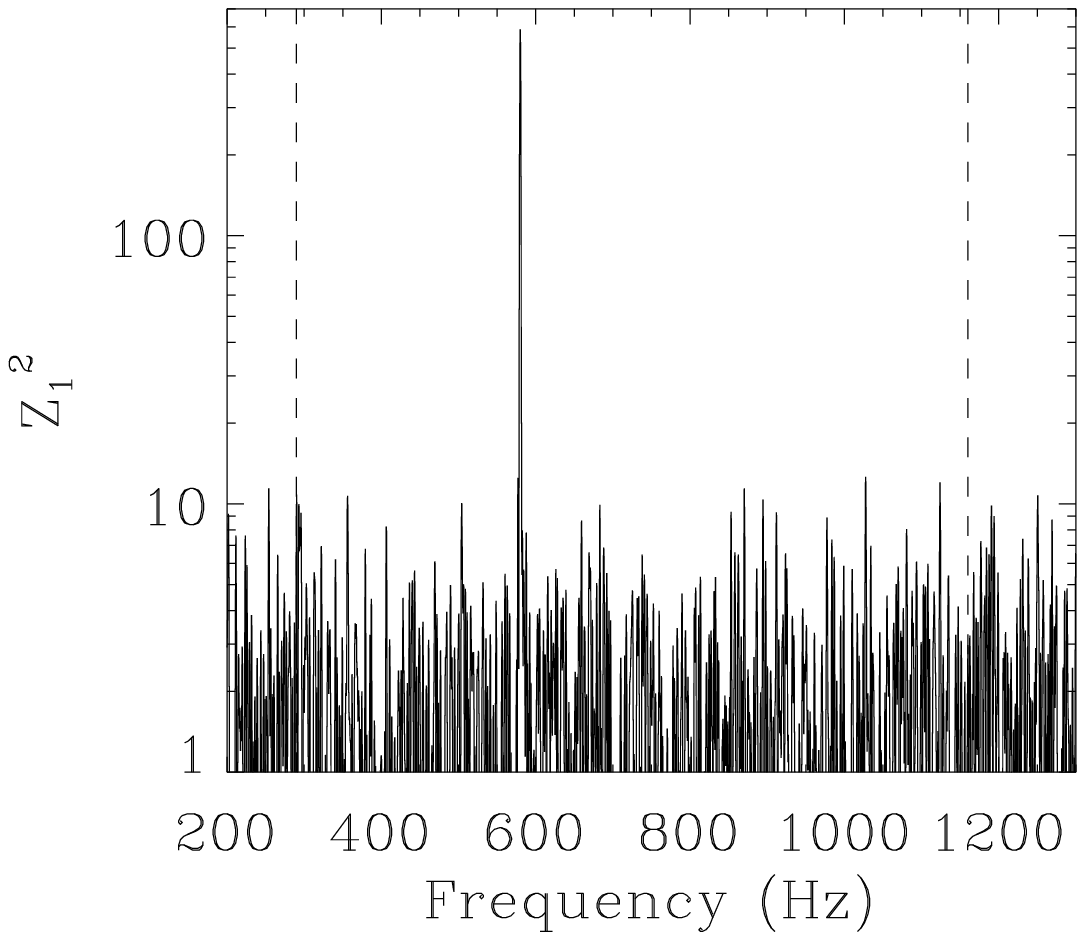}
\includegraphics[width=85mm]{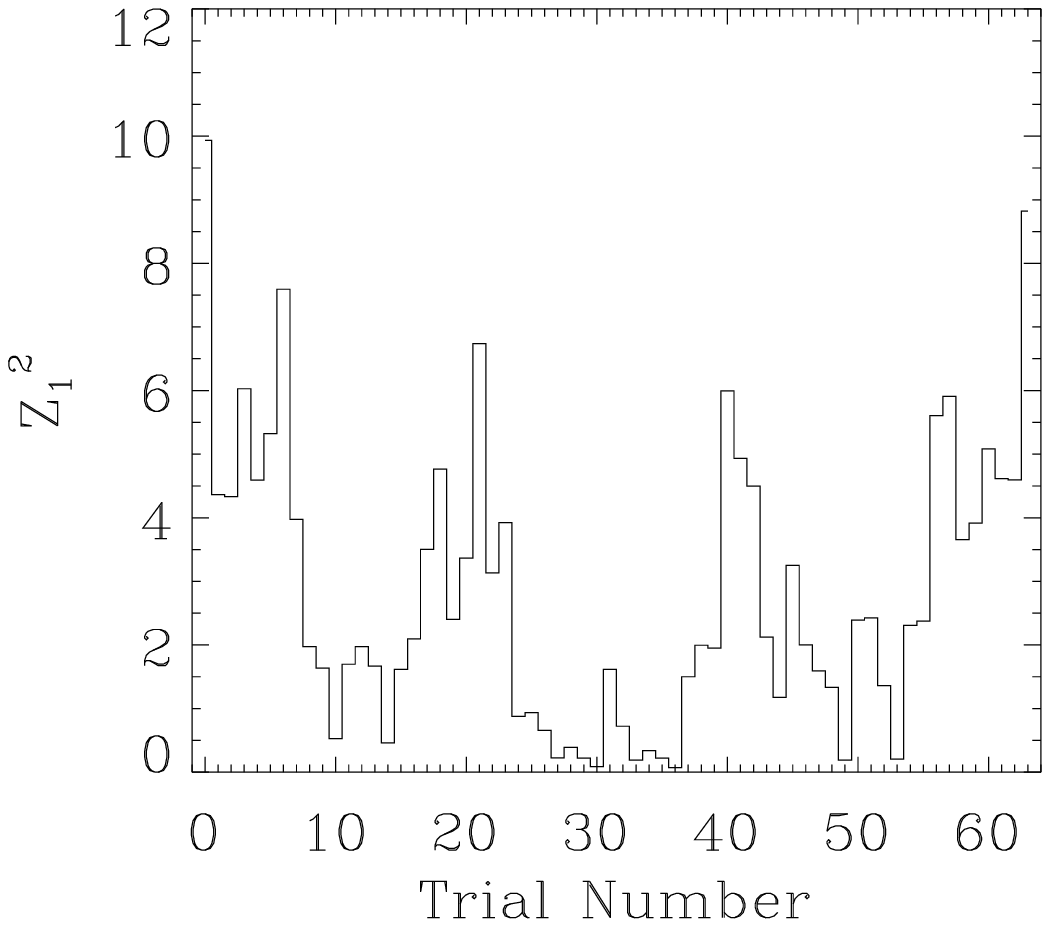}
\caption{Coherent $Z_{sum}^2$ for the 7 burst intervals (left). Note the 
logarithmic y-axis scale. The strong peak at 580 Hz has a power of $\sim 585$. 
The vertical dashed lines mark the
positions of the first subharmonic (290 Hz) and the first harmonic (1160 Hz),
neither of which are detected. The right panel shows $Z_{sum}^2$ for the 
64 possible phase permutations of the first subharmonic when combining 7 
independent segments, the largest power was $\sim 9.98$ and with 64 trials is 
not significant.}
\end{figure}

\section{SUMMARY AND CONCLUSIONS}

As outline above the inferences which can be made concerning neutron star 
structure and thermonuclear burning depend importantly on whether one or two 
hot spots produce the observed burst oscillations. So the question remains,
what is the spin frequency of 4U 1636-53, is it 290 Hz or 580 Hz? 
Based on the work presented here I am not able to confirm the 290 Hz 
interpretation presented by Miller (1999). In a further effort to try and 
confirm the subharmonic detection reported by Miller (1999), I used the $Z_1^2$
method outlined here to investigate the same set of bursts analysed by Miller 
(1999). I was able to coherently add the signals at 580 Hz, obtaining a peak 
$Z_1^2 = 135$, a value very close to that found by Miller (1999) using the 
``matched waveform filtering'' method, however, 
I did not find a significant peak at the 290 Hz subharmonic. I found a $Z_1^2$ 
of only $\sim 6.8$ which given 16 trials is not significant. Note, however, 
that an analysis using the $Z_1^2$ technique discussed here but applied to the 
{\it same} data intervals used by Miller (1999) confirms the significance of 
the signal at 290 Hz reported by Miller (C. Markwardt, personal communication).
It seems likely that the different data intervals selected by the two methods 
accounts for the different results. This conclusion is further supported by the
fact that I was able to increase the measured power at the 290 Hz subharmonic 
by including a larger interval from one of the bursts in the sample originally 
studied by Miller (1999) (ocurring on December 28, 1996 at 23:47:25 UTC: 
obsid 10088-01-07-02). In fact, it appears that the detection reported by 
Miller (1999), and confirmed by Markwardt (2001), is dependent strongly on this
one burst. Moreover, the inability to detect a subharmonic in subsequent bursts
reported here supports the notion that perhaps only a small fraction of bursts 
produce significant oscillations at 290 Hz. The results described here should 
not be seen as an attempt to discredit or disparage the previous work by Miller
(1999), rather, it is simply a matter of a result with very important 
implications requiring confirmation, especially in an independent data set. 

The attempt to detect the subharmonic in new data from 4U 1636-53 has led to
several interesting puzzles, the first of which is, why is the subharmonic 
apparently so weak most of the time? If two hot spots are indeed present, then 
the lack of a detection places a limit on the flux asymmetry from the two poles
as well as on how far from antipodal the spots may be. These new data suggest 
that the spots must be quite uniform, with the flux asymmetry being less than 
$\sim 13 \%$. With more detailed modelling it will be possible to convert the 
limit on the subharmonic amplitude into a constraint on how far apart the spots
have to be. If magnetic pooling is at work, then this could give a clue as to 
the magnetic field geometry on the neutron star. The lack of any harmonic 
structure is also very important. Both the stellar compactness (if small 
enough) and the relativistic rotational motion of the hot spot should introduce
harmonic structure into the pulse shape. The lack of harmonic structure 
suggests that the star cannot be bigger than some limit. As the compactness 
decreases the pulse becomes sharper and also more asymmetric because of the 
rotational aberration, thus detailed modelling and the limits derived here on 
the harmonic strength should lead to a lower bound on the compactness. 

The other intriguing result concerns the nature of the phase
shift indicated for burst 4. Although only one burst in this sample showed
evidence for a phase shift there are indications in other bursts from 4U 
1636-53 of similar behavior (Markwardt 2001). So far the putative spin 
down which is postulated to occur because
of the hydrostatic expansion of the thermonuclear burning layer has not been
definitively seen. This may be due to the fact that the expansion is 
expected to happen on a timescale shorter than the radiative diffusion time 
so that the shell has already puffed up before the photons have a chance to 
leak out and inform the world of the spin down. The phase shift in burst 4 
can be explained by a spin down during the interval when the oscillations fade,
and this might conceivably be evidence for the expansion induced shift, but 
naively one might expect that a more or less smooth spindown would occur, 
which is not what is seen. Since oscillations on the rising edge of bursts
typically fade as the flux increases, it is likely that either the spot or
spots are spreading during this time which will reduce the observed 
amplitude. The nature of this spreading is not well understood. It may be that
it could introduce a phase offset when a frequency derived 
during a time interval when spreading is significant is extrapolated to later
times when the spreading has stopped. 

\section*{ACKNOWLEDGEMENTS}

I would like to thank all the organizers of the COSPAR symposium on black holes
and neutron stars, and especially Andrzej Zdziarski, for putting together such 
a stimulating symposium and for being such gracious hosts. This work benefitted
greatly from discussions with Craig Markwardt, Jean Swank, Nitya Nath and 
Deepto Chakrabarty. I thank the referee, Cole Miller, for his insightful and 
helpful comments.

\end{document}